\documentclass{llncs}

\usepackage[utf8]{inputenc}
\usepackage[T1]{fontenc}
\usepackage[final]{graphicx}
\usepackage{multirow}
\usepackage{cite}
\usepackage{epstopdf}
\usepackage[hyphens]{url}
\usepackage{amsmath}
\usepackage{amssymb}
\usepackage{array}
\newcolumntype{C}[1]{>{\centering\arraybackslash$}p{#1}<{$}}
\usepackage[Symbol]{upgreek}
\usepackage{dsfont}
\usepackage{color,colortbl}
\definecolor{Gray}{gray}{0.9}
\usepackage{bigstrut}
\usepackage{enumerate}
\usepackage{setspace}

\usepackage{algorithm}
\usepackage{algpseudocode}

\usepackage[left=3cm,right=3cm,top=3cm,bottom=3cm]{geometry}

\usepackage{adjustbox}

 \usepackage[english]{babel}

\graphicspath{{img/}} 
\begin{document}

\title{Portfolio Risk Assessment using Copula Models\protect\footnote{%
    Submitted to the International Conference on Applied Research in Economics, Perm, Russia, 2017}}

\author{
Mikhail Semenov\inst{1} \and Daulet Smagulov\inst{2}}

\institute{
Tomsk Polytechnic University, Tomsk, Russia,\\
\inst{1}\email{sme@tpu.ru}, \
\inst{2}\email{master.daulet@gmail.com}}

\maketitle

\begin{abstract}
In the paper, we use and investigate copulas models to represent multivariate dependence in financial time series. We propose the algorithm of risk measure computation using copula models. Using the optimal mean-$CVaR$ portfolio we compute portfolio's Profit \& Loss series and corresponded risk measures curves.
Value-at-risk and Conditional-Value-at-risk curves were simulated by three copula models: full Gaussian, Student's~$t$ and regular vine copula. These risk curves are lower than historical values of the risk measures curve. All three models have superior prediction ability than a usual empirical method. Further directions of research are described.

\end{abstract}

\section{Introduction}

Common measures of risk  used in risk management are "value at-risk" ($VaR$) and "conditional value at-risk" ($CVaR$) which can be determined for different levels of significance. In the paper \cite{Kritski2007} it is shown that in the presence of a correlation of profits and losses series of assets, one-dimensional risks' measures $VaR$ and $CVaR$  assess the portfolio risk incorrectly. Therefore, to assess the portfolio risk, it is necessary to use $d$-dimensional measures determined through the multivariate dependence.

There are many different approaches that actively used in applications to represent multivariate dependence, for instance, principal component analysis, Bayesian networks, fuzzy techniques, factor analysis, and joint distribution function \cite{Huynh2014, Kole2007}. The dependence among the random variables $x_1, x_2, \ldots, x_d$, is completely described by the joint distribution function $F_X(x_1, x_2, \ldots, x_d)$. The idea of separating $F_X(x_1, x_2, \ldots, x_d)$ in two parts -- the one which describes the dependence structure, and the other one which describes the marginal behavior, leads to the concept of copula.
%
%
In 1959 A.~Sklar \cite{Sklar1959}
first proved the theorem that a collection of marginal
distributions can be coupled together via a \textit{copula} to form a
multivariate distribution. The copula contains all the information
about the dependence structure of the involved variables. In the paper \cite{Penikas2010} the author introduced copula-models' concepts and its application to the different financial issues including the task of risk measurement.

Many ways to describe financial data using Gaussian (normal) distribution exist today. It is well known \cite{Xu2008} that a full Gaussian copula, i.e., a Gaussian copula constructed by Gaussian (normal) marginals,
is another way to describe Markowitz's mean-variance portfolio
theory. On the other hand, a lot of empirical studies have shown that Gaussian distribution has a lot of problem in dependence description of financial data \cite{Rachev2005, Limp2011, Wilmott2007}. Moreover, in the standard \cite{EBA2015} is proposed that Gaussian copulas are not to be used for operational risk modelling. For instance, a Student's~$t$ copula with few integer degrees of freedom (three or four) in most cases appears more appropriate.
Closed-form expressions to calculate the sensitivity of the risk measure, $CVaR$, were proposed in the paper \cite{Stoyanov2013}.

The primary objective of this paper is to compare the financial risk measures in framework of portfolio management by deriving the relevant parameters of the copula models from prices of traded assets.



Studies \cite{Xu2008, Lourme2016, Ane2003, Kole2007} of price risks in framework of portfolio management in many respects are similar to each other and differ only in the used data and insignificant variations in the copula models estimation. Among set studies, we single out the paper by Ane~et~al \cite{Ane2003} that was one of the first where authors selected the dependence structure of international stock index returns through the Clayton copula.
Lourme~et~al \cite{Lourme2016} address the issue of testing the the full Gaussian
and Student's~$t$ copulas in a risk management framework. They proposed the $d$-dimensional compact Gaussian and Student's~$t$ confidence area inside of
which a random vector with uniform margins on
$(0, 1)$ falls with probability~$\alpha$.
The results evidence that the Student's~$t$ copula $VaR$ model is an attractive alternative to the Gaussian one.
A portfolio of stocks, bonds and real estate was considered \cite{Kole2007} to determine the importance of selecting the right copula for risk management. The Gaussian, the Student’s~$t$ and the Gumbel copulas to model the dependence of the daily returns on indexes that approximate these three asset classes were tested. Then with Value-at-Risk computations was established that the Gaussian copula is too optimistic on the diversification benefits of the assets, while the Gumbel copula is too pessimistic.

Estimation of the unknown parameters is an important problem.
At present, many algorithms for constructing copulas have been
designed.
For copula model
estimation, there exist three methods: the full parametric method
\cite{Patton2006}, the semiparametric method \cite{Chen2006, Lourme2016}, and
the nonparametric method \cite{Fermanian2003, Kim2007}. The full parametric
method is implemented via two-stage maximum likelihood estimation
(MLE) proposed by H.~Joe \cite{Joe1997, Joe2014}. 
The copula is fitted using the two-stage parametric MLE approach,
also referred to as the Inference Functions for Margins (IFM) method.
This method fits a copula in two steps:
    (1) estimate the parameters of the marginals, and
    (2) fix the marginal parameters to the values estimated in
first step, and subsequently estimate the copula parameters.

For the the bivariate case, the main families of copulas are: ellipse (Gaussian,
Student's~$t$), archimedean (Clayton, Frank, Joe), and extreme (Gumbel,
Cauchy).
In the dissertation research \cite{Xu2008}, the two-stage  MLE method
was applied, while the author uses all possible combinations of
different marginal distributions (Gaussian, the Student's $t$, and
skewed $t$ distribution) and different archimedian copulas in the
estimation and testing process. The decision of choosing the
marginal distribution is taken after the second step of MLE
method. For this purpose, a modification of the superior
predictive ability of the Hansen test \cite{Hansen2005} was
proposed; it allows one to identify a copula that has superior
forecasting ability.
%

 Multivariate copulas based on the one distribution (for instance, normal or Student's~$t$) or on one the generator function lack the flexibility of accurately modeling the dependence among larger numbers of variables \cite{Brechmann2013}. These lacks predetermined the direction of further research, as a result of which the regular vine copulas' (R-vine) concept was proposed by Joe \cite{Joe1996} and developed in more detail in \cite{Brechmann2013, Cooke2015}.
R-vine copulas are a flexible graphical model for describing multivariate copulas built up using
a cascade of bivariate copulas (two-dimensional function). This copula is easier to be interpreted and visualized, and we have a lot of methods to work with it today \cite{Cooke2015, Czado2010, Dissmann2013}.
For instance, in the study \cite{Dissmann2013} a novel algorithms for evaluating
a \emph{regular vine copula} parameters and simulating
from specified R-vines were proposed. The selection of the R-vine
tree structure based on a maximum spanning tree algorithm (MST),
where edge weights are chosen appropriately to reflect large
dependencies.

In this paper, we perform the computation of the copula models on the four time series of closing daily prices of futures. We use the two-stage maximum likelihood estimation by Joe \cite{Joe1997,Joe2014} for the copula models.

The contributions of this article are threefold. First, we show that full Gaussian and Student’s~$t$ as well as regular vine copula models can be used to represent multivariate dependence in short finance time series (253 observations only).  While the impact of copulas has been studied in relation to long time series \cite{Kole2007, Lourme2016, Dissmann2013}. Second, we use non-normal marginal distributions: the Hyperbolic \cite{Barndoff1983}, stable Paretian \cite{Rachev2005} and Meixner \cite{Schoutens2002} distributions, as the possible forms of marginal distributions. Stoyanov~et~al \cite{Stoyanov2013} address this issue, but only consider the symmetric stable Paretian, and the Student's~$t$, and generalized normal distributions. Third, constructing the regular vine copula model the we use 
non-integer degrees of freedom for two-parameter copulas that substantially extends the possibility of copulas' models in framework of portfolio risk management. 

%
%
In order to compare the performances of the different risk measures  ($VaR$, $CVaR$), we  propose to use the Monte-Carlo simulation on the mean-CVaR optimal portfolio.
The advantage of the $CVaR$ portfolio optimization is that one can formulate the mean-$CVaR$ portfolio optimization problem as a linear programming problem \cite{Rock2000}.
We plot and compare the historical Profit \& Loss series with $VaR$ and $CVaR$ curves at the different levels estimated through copula models.


The rest of the paper is arranged as follows. In Section~\ref{Methodology}, we introduce the methodology and
the dataset, on which our approach has been tested. 
In Section~3 we calculate future portfolios,
and examined the financial risk measures on the proposed portfolios.
Finally, the conclusion is included.

\section{Dataset and Methodology}\label{Methodology}

\subsection{Dataset}

In this research, we examine  time series of closing daily prices of stock futures for companies: MMC Norilsk Nickel PJSC (GMKR), Gazprom PJSC
(GAZP), Sberbank PJSC (SBER) respectively, and the future on RTS
index (RTS). Our sample covers the period of one year from to December 16, 2015, to
December 16, 2016. Denote them as GMKR, GAZP, SBRF and RTS respectively. All that data regarding the futures prices were collected from the Finam Holdings service (finam.ru). 


\subsection{Data Processing}
First, we have converted an initial data set to logarithmic returns (log-returns) so that we could use a stable data set that can be used for time series modeling and subsequent transformation. Equation~(\ref{log-returns}) transforms a price series $p$ into a log-returns $r$ series for each asset:

\begin{equation}\label{log-returns}
r_{t,i}=\log \frac{p_{t,i}}{p_{t-1,i}},
\end{equation}

\noindent where $i \in \overline{1, d}$, $d$ is the number of assets, $t\in \overline{1, T}$ is a time point, in our case $T=253$, $d = 4$.

Since we are going to deal with financial time-series, which has a nonlinear dependence,
we use rank correlation coefficients: Kendall's $\tau$ and Spearman's $\rho$. In further calculations, we used Kendall's~$\tau$ \cite{Dissmann2013}.
Correlation matrices are shown in Table~\ref{Cor1}.

\begin{table}
\centering
\caption{Spearman's $\rho$ and Kendall's $\tau$}
\label{Cor1}
\setlength{\tabcolsep}{4pt}
\begin{tabular}{l|rrrr|rrrr}
\hline
\multicolumn{1}{c}{} & \multicolumn{4}{c}{Spearman’s $\rho$} & \multicolumn{4}{c}{Kendall’s $\tau$} \\ \hline
log-returns & \multicolumn{1}{c}{RTS} & \multicolumn{1}{c}{SBRF} & \multicolumn{1}{c}{GAZP} & \multicolumn{1}{c|}{GMKR} & \multicolumn{1}{c}{RTS} & \multicolumn{1}{c}{SBRF} & \multicolumn{1}{c}{GAZP} & \multicolumn{1}{c}{GMKR} \\ \hline
RTS  & 1 & 0.78 & 0.67 & 0.27 & 1 & 0.58 & 0.49 & 0.18 \\
SBRF & 0.78 & 1 & 0.58 & 0.29 & 0.58 & 1 & 0.41 & 0.19 \\
GAZP & 0.67 & 0.58 & 1 & 0.35 & 0.49 & 0.41 & 1 & 0.24 \\
GMKR & 0.27 & 0.29 & 0.35 & 1 & 0.18 & 0.19 & 0.24 & 1 \\ \hline
\end{tabular}
\end{table}

\subsection{Estimate the parameters of the marginals}
Many ways to describe financial data using Gaussian (normal) distribution exist today \cite{Json1949}. On the other hand, a lot of empirical studies have shown that Gaussian distribution has a lot of problem in description of financial data \cite{Rachev2005, Limp2011, Wilmott2007}. Various non-normal distributions have been proposed for modeling extreme events, we choose the Hyperbolic \cite{Barndoff1983}, Stable \cite{Nolan2009, Rachev2005, Stoyanov2013} and Meixner \cite{Schoutens2002} distributions as the three possible forms of marginal distributions.
For the sake of brevity, the well-known formulae corresponding to the three possible forms of marginal distributions
are not reported here, but are available in \cite{Barndoff1983, Nolan2009, Rachev2005, Stoyanov2013, Schoutens2002}.

A hyperbolic distribution $H(\pi, \zeta, \delta, \mu)$ is four parameter distribution \cite{Barndoff1983} that determines with $\pi$ is the steepness of the distribution, $\zeta$ determines the symmetry, the distribution is symmetrical about the location parameter $\mu$ if $\zeta=0$, and $\delta$ is the scale parameter. 
%
A stable distribution $S(\alpha, \beta, \gamma, \mu)$ is described by four parameters \cite{Rachev2005, Nolan2009, Stoyanov2013}. The parameters and their meaning are: $\alpha$, which determines the tail weight or the distribution's and a kurtosis, $\beta$, which determines the distribution's skewness, $\gamma$ is a scale parameter, and $\mu$ is a location parameter. 
%
A Meixner distribution $M(\alpha, \beta, \delta, \mu)$ has four parameters: $\mu$ is the location parameter, $\alpha$ is the scale parameter, $\beta$ is the skewness parameter, and $\delta$ is the shape parameter \cite{Schoutens2002}. 

The first two parameters of mentioned above distributions 
are the most important as they identify two fundamental properties that are atypical of the normal distribution –– heavy tails and asymmetry \cite{Stoyanov2013}. Parameters for hyperbolic distribution have been estimated by the Nelder-Mead method, for the stable and the Meixner distribution –– by the Cramér–von Mises distance. The results obtained for the log-returns are shown in Table~\ref{dist-pars}. 

\begin{table}[t]
\centering
\caption{Parameter estimates for log-returns marginal
distribution models}
\label{dist-pars}
\setlength{\tabcolsep}{4pt}
\begin{tabular}{lcrrrr}
\hline \multicolumn{2}{c}{Parameters} & \multicolumn{1}{c}{RTS} &
\multicolumn{1}{c}{SBRF} & \multicolumn{1}{c}{GAZP} &
\multicolumn{1}{c}{GMKR} \bigstrut \\ \hline
Hyperbolic  &    $\pi$ & $-$0.08291 &    0.12785 &    0.06765 &    0.24050 \\
            &  $\zeta$ &    1.83534 &    0.00317 &    0.67834 &    0.01527 \\
            & $\delta$ &    0.01918 &    0.00004 &    0.00618 &    0.00017 \\
            &    $\mu$ &    0.00445 & $-$0.00123 & $-$0.00090 & $-$0.00479 \\ \hline
Stable      & $\alpha$ &    1.86195 &    1.80091 &    1.85290 &    1.75151 \\
            &  $\beta$ &    0.19345 &    0.90962 &    0.84324 &    0.87382 \\
            & $\gamma$ &    0.01236 &    0.01019 &    0.00816 &    0.00920 \\
            & $\delta$ &    0.00160 &    0.00072 & $-$0.00059 & $-$0.00153 \\ \hline
Meixner     & $\alpha$ &    0.01925 &    0.02726 &    0.02209 &    0.00464 \\
            &  $\beta$ &    0.47214 &    0.68221 &    0.26678 &    2.12144 \\
            & $\delta$ &    1.74613 &    0.67245 &    0.69809 &    4.22827 \\
            &    $\mu$ & $-$0.00575 & $-$0.00345 & $-$0.00158 & $-$0.03413 \\ \hline
\end{tabular}
\end{table}

\begin{figure} 
  \centering
  \includegraphics[height=0.93\textheight]{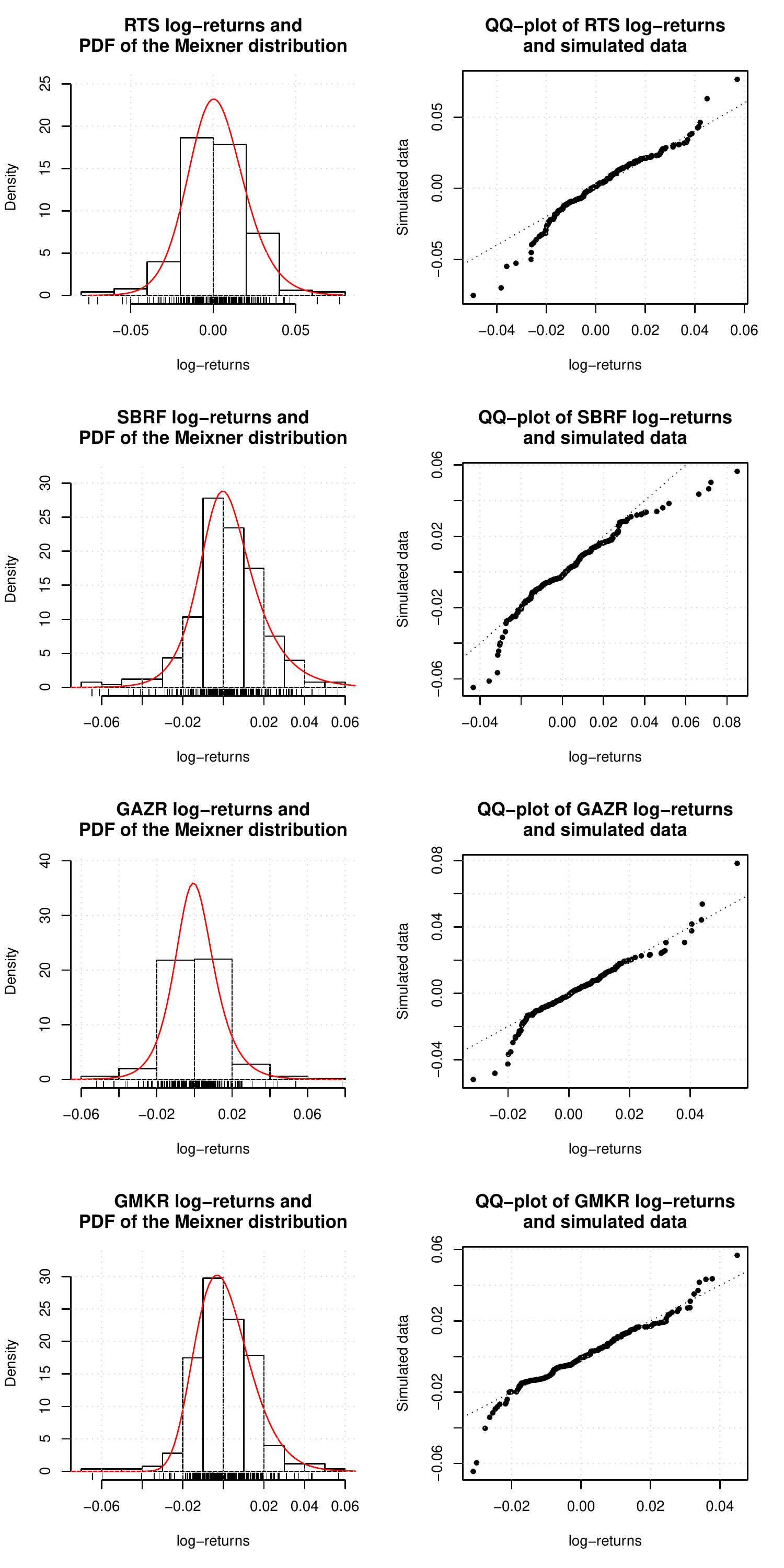}%
  \caption{Histogram (left) and $Q$-$Q$ plot (right) of the observations and selected marginal distribution}
  \label{fig1}
\end{figure}

We computed the Kolmogorov–Smirnov (KS) test, Anderson–Darling (AD) test and Cramér–von Mises (CvM) test between the
empirical marginals and the fitted marginals,  the associated $p$-values are reported in Table~\ref{dist-test}, respectively.
Since in all the cases the $p$-values are quite high, it indicates that the proposed marginal distributions are indeed a good model for data set. According to results of the CvM test, we take the Mexiner distribution as the marginal for each underlying asset (denoted by grey in Table~\ref{dist-test}). Fig.~\ref{fig1} shows comparison of each observations set and corresponding distribution on the histogram and the $Q$-$Q$ plot, respectively.

\begin{table}[t]
\centering
\caption{The associated $p$-values of statistical tests of marginal distribution parameters}
\label{dist-test}
\setlength{\tabcolsep}{4pt}
\begin{tabular}{llcccc} \hline
\multicolumn{1}{c}{Test} & \multicolumn{1}{c}{Marginal} & RTS & SBRF & GAZP & GMKR \bigstrut \\ \hline
\multirow{3}{2.3cm}{Kolmogorov--Smirnov}& Hyperbolic    & 0.73 & 0.14 & 0.96 & 0.25 \\
                                        & Stable        & 0.89 & 0.83 & 0.98 & 0.96 \\
                                        & Meixner       & 0.92 & 0.86 & 1.00 & 0.97 \\ \hline
\multirow{3}{2cm}{Anderson--Darling}    & Hyperbolic    & 0.69 & 0.39 & 0.98 & 0.22 \\
                                        & Stable        & 0.70 & 0.26 & 0.72 & 0.73 \\
                                        & Meixner       & 0.49 & 0.52 & 0.98 & 0.00 \\ \hline
\multirow{3}{1.8cm}{Cramér--von~Mises}  & Hyperbolic    & 0.65 & 0.34 & 0.98 & 0.22 \\
                                        & Stable        & 0.78 & 0.73 & 0.99 & 0.89 \\
                                        & Meixner       & \cellcolor{Gray}0.81 & \cellcolor{Gray}0.79 & \cellcolor{Gray}1.00 & \cellcolor{Gray}0.95 \\ \hline
\end{tabular}
\end{table}

\begin{figure} 
  \centering
  \includegraphics[width=0.49\textwidth]{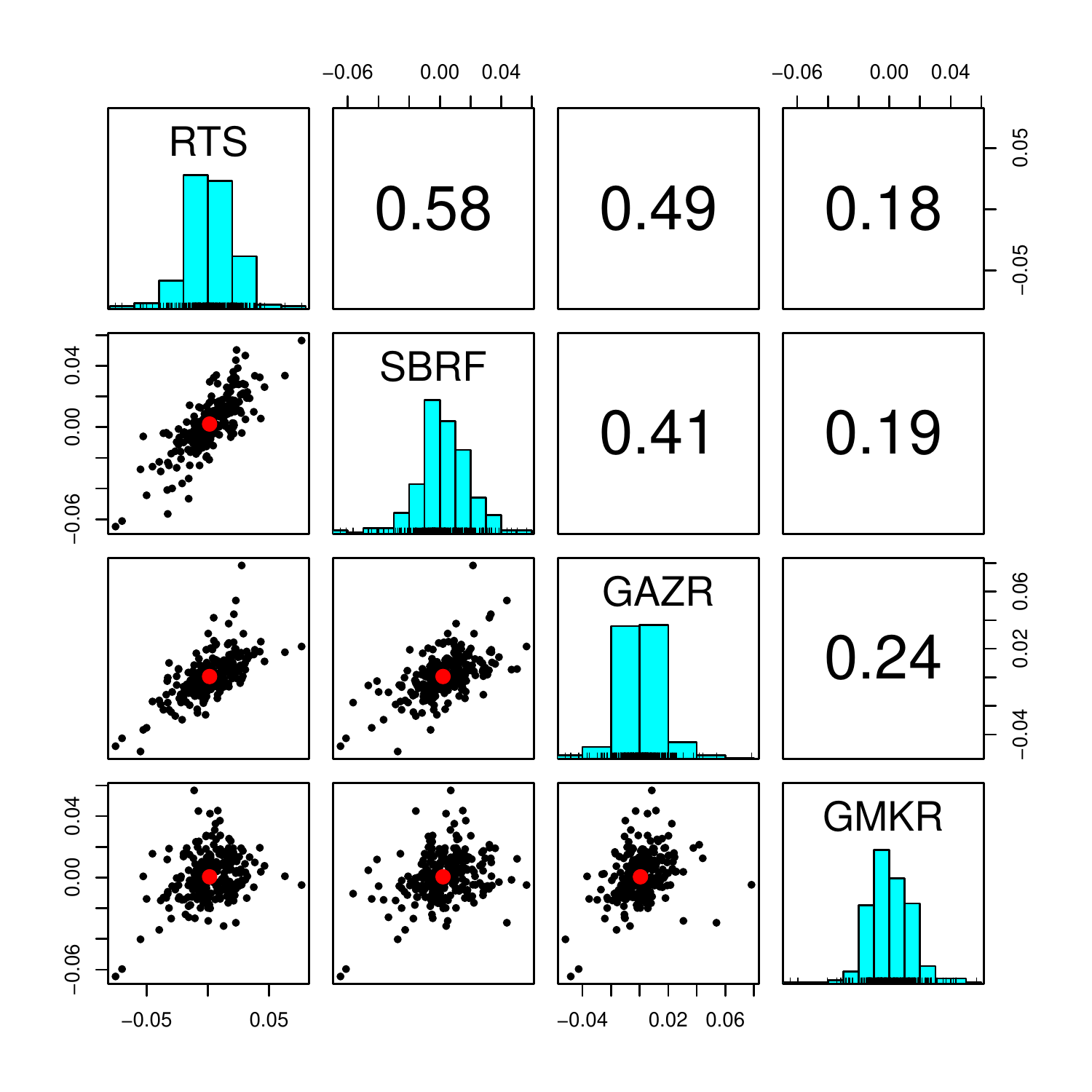}
  \includegraphics[width=0.49\textwidth]{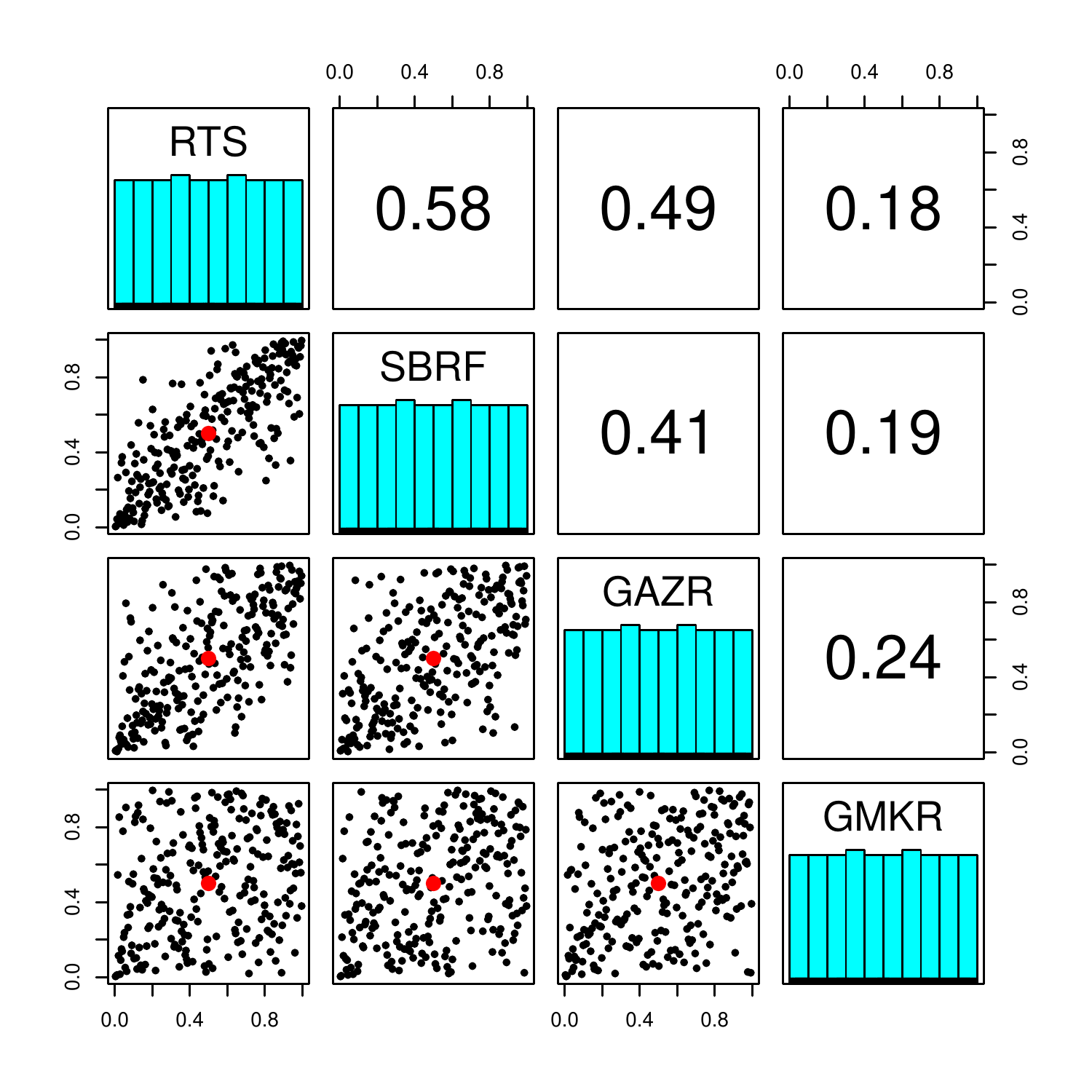}
  \caption{Historical observations (left) and pseudo-observations (right). Upper  triangular  matrix:  Kendall’s $\tau$ values between pairs. Lower triangular matrix: Bivariate scatter plots. Diagonal: histograms of marginals.}
  \label{pairs}
\end{figure}


\subsection{Estimate the copula parameters}


At this stage, we suggest the constructing of copula using two type of copula models: multivariate copula and regular vine (R-vine) copula. For the sake of brevity, the known formulae corresponding to the 
copulas are not reported here, but are available in \cite{Nelsen1999, Joe2014,Czado2010,Cooke2015}.

First, we should generate points of the empirical copula, 
called as \textit{pseudo-observations}. 
Considering  Eq.~(\ref{log-returns}) $\boldsymbol{r}_i = (r_{1,i}, \ldots, r_{T,i})^\intercal$ for all historical observations (log-returns) $i \in \overline{1,d}$, then pseudo-observations are defined via the Eq.~(\ref{pobs}):

\begin{equation} \label{pobs}
    u_{t,i} = \frac{\text{rg}(r_{t,i})}{T + 1},\ \forall \ t \in \overline{1,T},\ i \in \overline{1,d},
\end{equation}

\noindent where $\text{rg}(r_{t,i})$ denotes the rank of $r_{t,i}$ (from lowest to highest) of the observed values $r_{\tau,i}, \tau \in \overline{1,T}$ \cite{Copula}. Each element $u_{t,i}$ is
between $0$ and $1$. Pairs plots of the joint distribution of observed data and the pseudo-observations are shown on Fig.~\ref{pairs}.


In this study we use elliptical copulas of two families: Gaussian (normal) and Student's~$t$ copulas.
To estimate the copula parameters we use 
pseudo-observations~Eq.~(\ref{pobs}). The decision of choosing the copula parameters is taken by <<Inversion of Kendall’s tau>> method \cite{Koj2010}. Then we execute a parametric bootstrap-based goodness-of-fit (GoF) test of elliptical copulas to check their quality \cite{Gen2009}. Estimated parameters and the test results are shown in Table~\ref{CopPars}. As one can see the parameters of elliptical copulas and results of GoF test are very close each to other.


The negative side of using multivariate copula model is that we can not (a) check the quality, and (b) construct the cumulative distribution function of Student's~$t$ copula with non-integer degrees of freedom.
Alternative way to construct copula models is using R-vine copulas. As we know from \cite{Bedfort2002}, a $d$-dimensional vine is a copula constructed of $d(d - 1)/2$ usual bivariate copulas. 
%
The main advantage of this type of models is that each component copula of vine represents a pair-copula (two-dimensional function). 
This copula is easier to be interpreted and visualized, and we have a lot of methods to work with it today \cite{Cooke2015, Czado2010, Dissmann2013}.

Following the study \cite{Dissmann2013} we use absolute empirical Kendall's $\tau$ as a measure of dependence, since it makes it independently of the assumed distribution. 
We use the same method \cite{Koj2010} to estimate the parameters as we did it with multivariate copulas. Also, we can use non-integer degrees of freedom for copulas with two parameters. 
In addition, we choose different families for each pair \cite{Vuong1989, Clarke2007, Bel2010}. 
Using abbreviations for copula types: $SG$~-- Survival Gumbel, $SC$~-- Survival Clayton, $SBB1$~-- Survival Clayton-Gumbel, $BB7$~-- Joe-Clayton, $Ind$~-- independence copula, the estimated R-vine copula is given by \cite{Czado2010}:
\begin{gather} \label{Tree}
    M = \left(
        \begin{array}{*4{C{1em}}}
        2 &   &   &   \\
        4 & 1 &   &   \\
        3 & 4 & 3 &   \\
        1 & 3 & 4 & 4 \\
        \end{array} \right), \ 
    T = \left(
        \begin{array}{llll}
        SC\ (0.116) &  &    \\
        BB7\ (1.084, 0.058) &\ Ind\  &   \\
        SBB1\ (0.229, 2.005)\ &\ SG\ (1.935)\ &\ SG\ (1.336)\ \\
        \end{array} \right),
\end{gather}
where $M$ is the matrix of the R-vine array structure, numbers 1, 2, 3, and 4 denote the log-returns of RTS, SBRF, GAZP, GMKR series respectively, $T$ is the matrix containing information about family and parameters of each bivariate (one- or two parameter) copula of R-vine. 

\begin{table}[t]
\centering
\caption{The parameters estimation for multivariate Gaussian, Student's~$t$, and R-vine copulas}
\label{CopPars}
\setlength{\tabcolsep}{8pt}
\begin{tabular}{l|r|rr}
\hline
\multicolumn{1}{c}{\multirow{2}{*}{Copula}} & \multicolumn{1}{c}{\multirow{2}{*}{Parameters}} & \multicolumn{2}{c}{GoF test results} \bigstrut \\ \cline{3-4}
\multicolumn{1}{c}{} & \multicolumn{1}{c}{} & \multicolumn{1}{c}{Statistic} & \multicolumn{1}{c}{$p$-value} \bigstrut \\ \hline
Gaussian      & $\rho=0.5$           & $S_n=0.11$  & 0.00495  \bigstrut \\
Student's~$t$ & $\rho=0.5,\ \nu=4$   & $S_n=0.13$  & 0.00495  \bigstrut \\
R-vine        & see Eq.~(\ref{Tree}) & $W=15.15$   & 0.95    \bigstrut \\ \hline
\multicolumn{4}{l}{$\rho$ –- the correlation copula parameter,} \bigstrut[t] \\
\multicolumn{4}{l}{$\nu$ –- degrees of freedom of Student's~$t$ copula.} \bigstrut[b]
\end{tabular}
\end{table}

To check the estimated parameters, we use a goodness fit tests based on the Cramer-von Mises statistic, $S_n$,  \cite{Koj2010} and the White’s information matrix equality, $W$, \cite{White1982}.
The result of test implementation is shown in Table~\ref{CopPars}. As one can see the $p$-values of elliptical copulas are less than corresponding $p$-value of the R-vine copula.



\section{Portfolio Application}

In this section, we present some simulation results
to compare the performances of
 the Value-at-Risk ($VaR$) and the Conditional Value-at-Risk ($CVaR$) on an equally weighted portfolio composed of $d=4$ assets. Then we applied the above results to compute the optimal weights of each asset, which is one of the major concerns in the field of portfolio risk management.

\subsection{Mean-Conditional-Value-at-Risk Portfolio Optimization}

The advantage of the $CVaR$ portfolio optimization is that we can formulate the mean-$CVaR$ portfolio optimization problem as a linear programming problem \cite{Rock2000}. 
If we can find a portfolio with a low $CVaR$, then it will also have a low $VaR$ \cite{wnc04}.
%
%
%
We assume a "full investment" portfolio with only long positions, furthermore, to avoid a corner portfolio case, let the minimal weight be limited by $0.05$. The mean-$CVaR$ portfolio weights we obtained are $0.05$, $0.05$, $0.4058$, and $0.4942$ for
RTS, SBRF,  GAZP, and GMKR respectively.


Now let us compare $VaR$ and $CVaR$ of equally weighted and mean-CVaR optimal portfolio obtaining for historical scenario by empirical methods \cite{Rock2000}. 
Take $\alpha = \{99.9\%,\, 99.5\%,\, 99\%,\, 95\%,\, 90\%\}$ as a level for $VaR$ and $CVaR$ computation. It is clear from the simulation results (Table~\ref{comparison}) that values of risk measures for optimal portfolio are distinctively better, as expected. We will considering further the optimal portfolio only.

\begin{table}
\centering
\caption{Risk measures and associated bias for different portfolios and level, $\alpha$}
\label{comparison}
\setlength{\tabcolsep}{5pt}
\begin{tabular}{c *{3}{|l@{\,/\,}l}} \hline
\multirow{2}{*}{Level, \%} & \multicolumn{6}{c}{$VaR_\alpha / CVaR_\alpha, \times 10^{-2}$} \bigstrut \\ \cline{2-7}
& \multicolumn{2}{c|}{Optimal} & \multicolumn{2}{c|}{Equiweighted} & \multicolumn{2}{c}{Bias $\times 10^{-2}$} \bigstrut \\ \hline
99.9 & $-5.70$ & $-5.83$ & $-6.17$ & $-6.29$ & $-0.47$ & $-0.46$ \bigstrut[t] \\
99.5 & $-5.13$ & $-5.58$ & $-5.45$ & $-6.05$ & $-0.32$ & $-0.48$ \\
99.0   & $-3.60$ & $-5.24$ & $-4.06$ & $-5.50$ & $-0.46$ & $-0.26$ \\
95.0   & $-1.77$ & $-2.87$ & $-2.02$ & $-3.30$ & $-0.25$ & $-0.44$ \\
90.0   & $-1.19$ & $-2.17$ & $-1.32$ & $-2.45$ & $-0.12$ & $-0.28$ \bigstrut[b] \\ \hline
\end{tabular}
\end{table}

\subsection{Efficient Algorithm of Risk Measure Computation using Copula Models}

Now we propose the following algorithm based on Monte-Carlo simulation of pseudo-observations to compute the risk measures. Here we use the estimation results of the copula models (Section~\ref{Methodology}, Tables~\ref{dist-pars},~\ref{CopPars}).

Algorithm~\ref{PnL-quantiles} represents the method we used to compute $VaR$ and $CVaR$. Method is based on Monte-Carlo simulation of pseudo-observations using proposed copula models with estimated parameters. In order to generate random samples (line \ref{Alg:simulation}) we used Gaussian and Student's $t$ copula parameters (Table~\ref{CopPars}) and the R-vine array structure, Eq.~(\ref{Tree}).  Then we transform each univariate pseudo-observation series to quantiles: $[0,1] \to \mathbb{R}$  (lines~\ref{Alg:transform:start}--\ref{Alg:transform:end}). In order to implement the transformation we take a quantile of each log-returns series using simulated pseudo-observations throw all dimensions of the copula as probabilities (line~\ref{Alg:transform}). Using optimal mean-$CVaR$ portfolio weights we compute portfolio's Profit \& Loss series (line~\ref{Alg:PnL}) and risk measures (line~\ref{Alg:risk-measures}).

The obtained results for VaR and CVaR are shown in Table~\ref{VaR-results} and \ref{ES-results} respectively. As we can see, we have all negative values of bias for vine copula. It means, that we will not loose more than we predict by this model. Thus, we can say that the R-vine copula model has superior 
forecasting ability than the Gaussian and the Student's~$t$ one.

\begin{algorithm}
\caption{Computation of Risk Measures by a Copula}
\label{PnL-quantiles}
\begin{algorithmic}[1]
    \Require Log-returns $\{r_{i,t}\}$, weights $w_i$ of optimal portfolio, $i \in \overline{1,d}$, $d$-dimensional copula c.d.f. with parameters and the array structure for R-vine, level $\alpha$ for $VaR_\alpha$ and $CVaR_\alpha$ calculation.
    \State Generate a sample of pseudo-observations $\{\hat{u}_{j,s}\} \in [0, 1]^d, \ j \in \overline{1,d}, \ s \in \overline{1, S}$ according to given copula.\label{Alg:simulation}
    \State Transform simulated pseudo-observations to univariate quantiles:
    \label{Alg:transform:start}
    \State Let $k = 1,2,\ldots,K,$ where $K = S \cdot d$.
    \For {$i \in \overline{1,d}$}
        \For {$j \in \overline{1,d}$}
            \For {$s \in \overline{1,S}$}
                \State Set $\hat{s}_{i,k} \gets \mathbb{Q}_{\hat{u}_{j,s}}  (r_i)$. \label{Alg:transform}
            \EndFor
        \EndFor
    \EndFor \label{Alg:transform:end}
    \State Compute the portfolio Profit \& Loss series:
    \For {$k \in \overline{1,K}$}
    \State $P\&L_k = \sum_{i=1}^d \hat{s}_{i,k} \cdot w_i$. \label{Alg:PnL}
    \EndFor
    \State Calculate $VaR_\alpha$, $CVaR_\alpha$ of Profit \& Loss series \label{Alg:risk-measures} 
    \Ensure $VaR_\alpha$ and $CVaR_\alpha$ of simulated Profit \& Loss series.
\end{algorithmic}
\end{algorithm}

\begin{table}[htbp]
\centering
\caption{$VaR$ obtained empirically and estimated by Gaussian\,/\,Student's~$t$\,/\,Vine copulas}
\label{VaR-results}
\setlength{\tabcolsep}{5pt}
\begin{tabular}{c|c|r@{\,/\,}r@{\,/\,}r|r@{\,/\,}r@{\,/\,}r} \hline
\multicolumn{1}{c|}{Level, \%} & \multicolumn{1}{c|}{Empirical, $\times 10^{-2}$} & \multicolumn{3}{c|}{Simulated, $\times 10^{-2}$} & \multicolumn{3}{c}{Bias, $\times 10^{-3}$} \bigstrut \\ \hline
99.9 & $-5.70$ & $-5.88$ & $-5.78$ & $-5.93$ & $-1.78$ & $-0.77$ & $-2.29$ \bigstrut[t] \\
99.5 & $-5.13$ & $-5.12$ & $-4.73$ & $-5.60$ & $0.15$ & $4.01$ & $-4.73$ \\
99.0   & $-3.60$ & $-3.85$ & $-4.07$ & $-4.10$ & $-2.53$ & $-4.74$ & $-4.96$ \\
95.0   & $-1.77$ & $-2.15$ & $-2.19$ & $-2.14$ & $-3.83$ & $-4.25$ & $-3.74$ \\
90.0   & $-1.19$ & $-1.46$ & $-1.50$ & $-1.50$ & $-2.66$ & $-3.06$ & $-3.05$ \bigstrut[b] \\ \hline
\end{tabular}
\end{table}

\begin{table}[htbp]
\centering
\caption{$CVaR$ obtained empirically and estimated by Gaussian\,/\,Student's~$t$\,/\,Vine copula}
\label{ES-results}
\setlength{\tabcolsep}{5pt}
\begin{tabular}{c|c|r@{\,/\,}r@{\,/\,}r|r@{\,/\,}r@{\,/\,}r} \hline
\multicolumn{1}{c|}{Level, \%} & \multicolumn{1}{c|}{Empirical, $\times 10^{-2}$} & \multicolumn{3}{c|}{Simulated, $\times 10^{-2}$} & \multicolumn{3}{c}{Bias, $\times 10^{-3}$} \bigstrut \\ \hline
99.9 & $-5.83$ & $-5.98$ & $-5.90$ & $-5.96$ & $-1.50$ & $-0.70$ & $-1.30$ \bigstrut[t] \\
99.5 & $-5.58$ & $-5.63$ & $-5.49$ & $-5.81$ & $-0.49$ & $0.89$ & $-2.36$ \\
99.0   & $-5.24$ & $-5.09$ & $-4.93$ & $-5.32$ & $1.51$ & $3.05$ & $-0.79$ \\
95.0   & $-2.87$ & $-3.27$ & $-3.31$ & $-3.32$ & $-4.03$ & $-4.41$ & $-4.54$ \\
90.0   & $-2.17$ & $-2.50$ & $-2.53$ & $-2.55$ & $-3.28$ & $-3.61$ & $-3.74$ \bigstrut[b] \\ \hline
\end{tabular}
\end{table}

\subsection{Stability Study and Risk Measure Curve}

\begin{figure} 
  \centering
  \includegraphics[width=\textwidth]{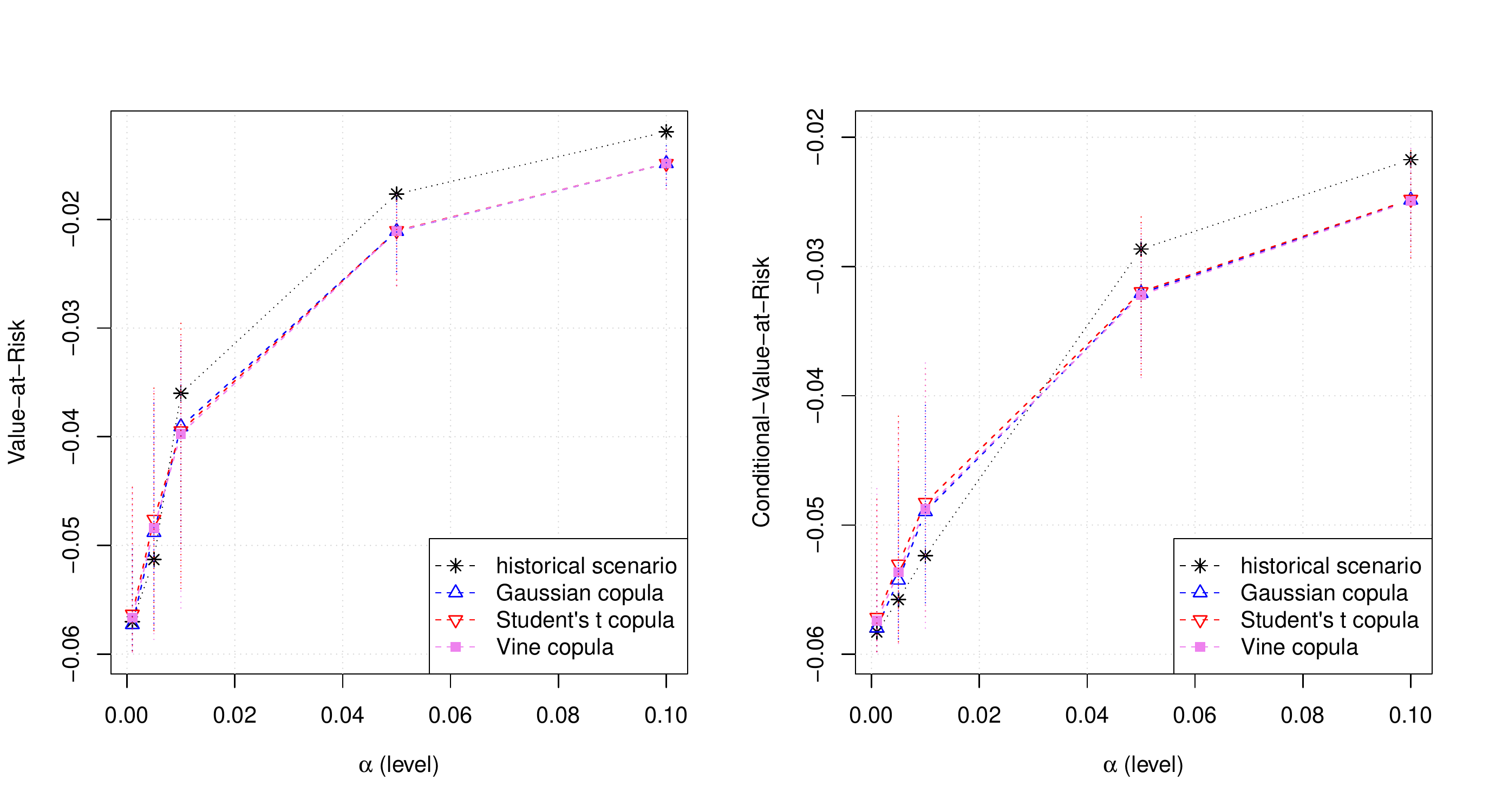}
  \caption{$VaR$ (left) and $CVaR$ (right). Historical values and bootstrap estimated (with 95\% confidence interval) by Gaussian\,/\,Student's~$t$\,/\,Vine copula}
  \label{VaR-ES.boot}
\end{figure}

Now let us make a stability research of proposed method. For this, we make bootstrap procedure replicating Algorithm~\ref{PnL-quantiles}. 
The obtained results are shown in Table~\ref{VaR-boot}, \ref{ES-boot} and illustrated on Fig.~\ref{VaR-ES.boot}.

We report the bias, the standard deviation (SD) and the mean square error (MSE) based  on  $N=200$  replications.  The bias value is better at lower levels: 99\%, 95\%, 90\% for $VaR$, and 95\%, 90\% for $CVaR$.
The SD and the MSE metrics show the greater instability of vine copula related to Gaussian (the most stable one) and Student's~$t$ model.

\begin{table}
\centering
\caption{VaR estimation by Gaussian\,/\,Student's~$t$\,/\,Vine~copula obtained by bootstrap procedure}
\label{VaR-boot}
\setlength{\tabcolsep}{5pt}
\begin{adjustbox}{max width=\textwidth}
\begin{tabular}{c*{4}{|r@{/}r@{/}r}} \hline 
\multicolumn{1}{c|}{Level, \%} & \multicolumn{3}{c|}{$\overline{VaR}_\alpha, \times 10^{-2}$} & \multicolumn{3}{c|}{Bias, $\times 10^{-3}$} & \multicolumn{3}{c|}{SD, $\times 10^{-2}$} & \multicolumn{3}{c}{MSE, $\times 10^{-6}$} \bigstrut \\ \hline
99.9 & $-5.74$ & $-5.72$ & $-5.59$ & $-0.34$ & $-0.22$ & $1.14$ & $2.83$ & $2.81$ & $4.94$ & $8.06$ & $7.88$ & $25.58$ \bigstrut[t] \\
99.5 & $-4.92$ & $-4.89$ & $-4.85$ & $2.11$ & $2.36$ & $2.83$ & $6.50$ & $6.86$ & $7.73$ & $46.54$ & $52.47$ & $67.42$ \\
99.0   & $-3.93$ & $-3.97$ & $-3.97$ & $-3.30$ & $-3.70$ & $-3.74$ & $5.21$ & $5.72$ & $6.86$ & $37.87$ & $46.27$ & $60.80$ \\
95.0   & $-2.09$ & $-2.13$ & $-2.13$ & $-3.19$ & $-3.59$ & $-3.67$ & $1.85$ & $2.22$ & $2.34$ & $13.55$ & $17.80$ & $18.93$ \\
90.0   & $-1.48$ & $-1.49$ & $-1.50$ & $-2.85$ & $-3.01$ & $-3.07$ & $0.94$ & $1.06$ & $1.17$ & $9.02$ & $10.14$ & $10.79$ \bigstrut[b] \\ \hline
\end{tabular}
\end{adjustbox}
\end{table}

\begin{table}
\centering
\caption{$CVaR$ estimation by Gaussian\,/\,Student's~$t$\,/\,Vine~copula obtained by bootstrap procedure}
\label{ES-boot}
\setlength{\tabcolsep}{5pt}
\begin{adjustbox}{max width=\textwidth}
\begin{tabular}{c*{4}{|r@{/}r@{/}r}} \hline 
\multicolumn{1}{c|}{Level, \%} & \multicolumn{3}{c|}{$\overline{CVaR}_\alpha, \times 10^{-2}$} & \multicolumn{3}{c|}{Bias, $\times 10^{-3}$} & \multicolumn{3}{c|}{SD, $\times 10^{-2}$} & \multicolumn{3}{c}{MSE, $\times 10^{-6}$} \bigstrut \\ \hline
99.9 & $-5.81$ & $-5.79$ & $-5.68$ & $0.18$ & $0.41$ & $1.49$ & $1.76$ & $2.24$ & $3.94$ & $3.13$ & $5.18$ & $17.65$ \bigstrut[t] \\
99.5 & $-5.45$ & $-5.42$ & $-5.32$ & $1.28$ & $1.63$ & $2.60$ & $3.87$ & $4.19$ & $5.67$ & $16.51$ & $20.09$ & $38.75$ \\
99.0   & $-4.92$ & $-4.91$ & $-4.86$ & $3.12$ & $3.25$ & $3.76$ & $4.42$ & $4.84$ & $6.08$ & $29.13$ & $33.90$ & $50.96$ \\
95.0   & $-3.20$ & $-3.24$ & $-3.22$ & $-3.39$ & $-3.70$ & $-3.56$ & $2.73$ & $3.10$ & $3.55$ & $18.92$ & $23.21$ & $25.22$ \\
90.0  & $-2.48$ & $-2.51$ & $-2.50$ & $-3.05$ & $-3.32$ & $-3.27$ & $1.90$ & $2.17$ & $2.45$ & $12.90$ & $15.71$ & $16.62$ \bigstrut[b] \\ \hline
\end{tabular}
\end{adjustbox}
\end{table}



\begin{figure} 
  \centering
  \includegraphics[width=\textwidth]{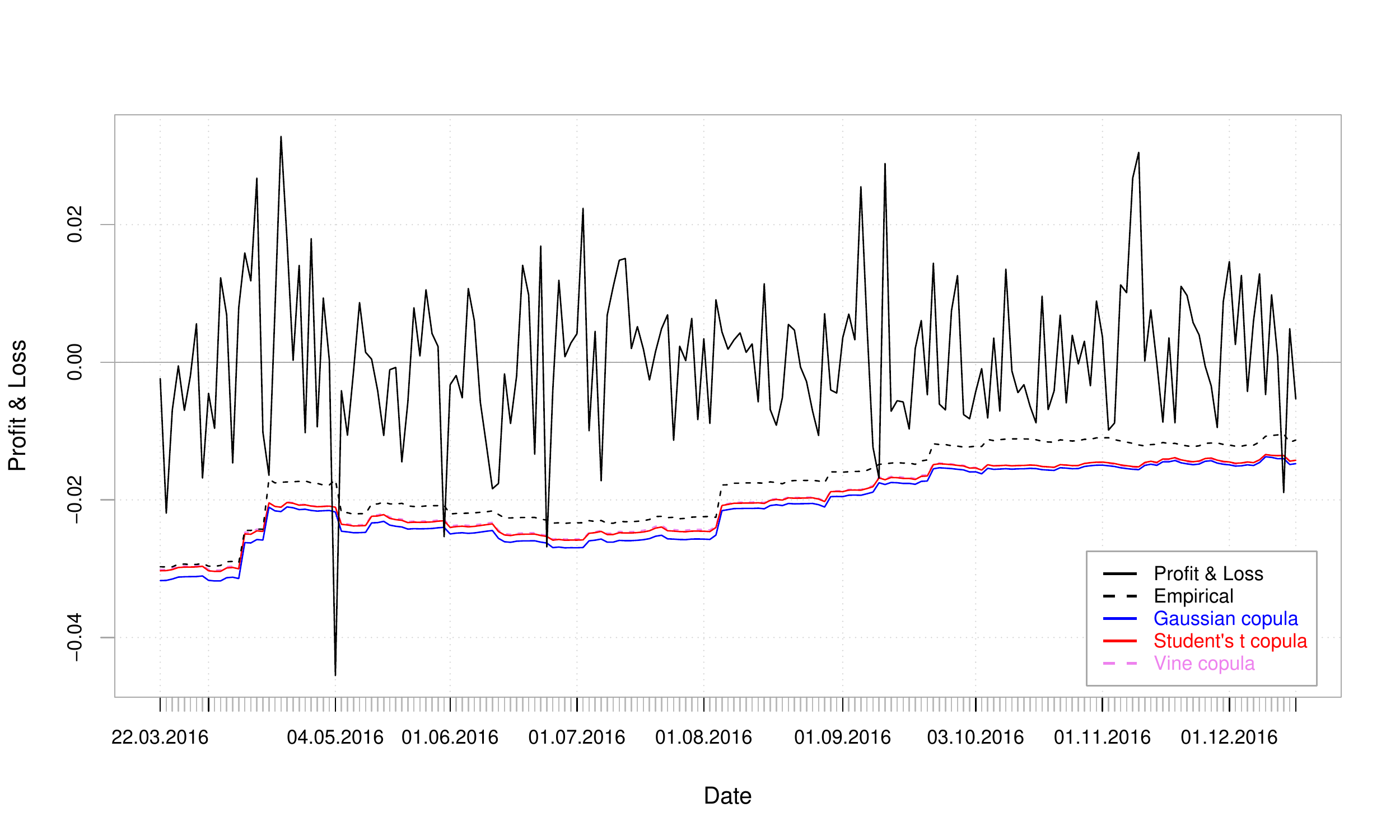}
  \caption{Profit \& Loss (in black), $VaR$ curve (at the 95\% level) obtained empirically and estimated by Gaussian\,/\,Student's~$t$\,/\,Vine copula}
  \label{VaR-curve}
\end{figure}

\begin{figure} 
  \centering
  \includegraphics[width=\textwidth]{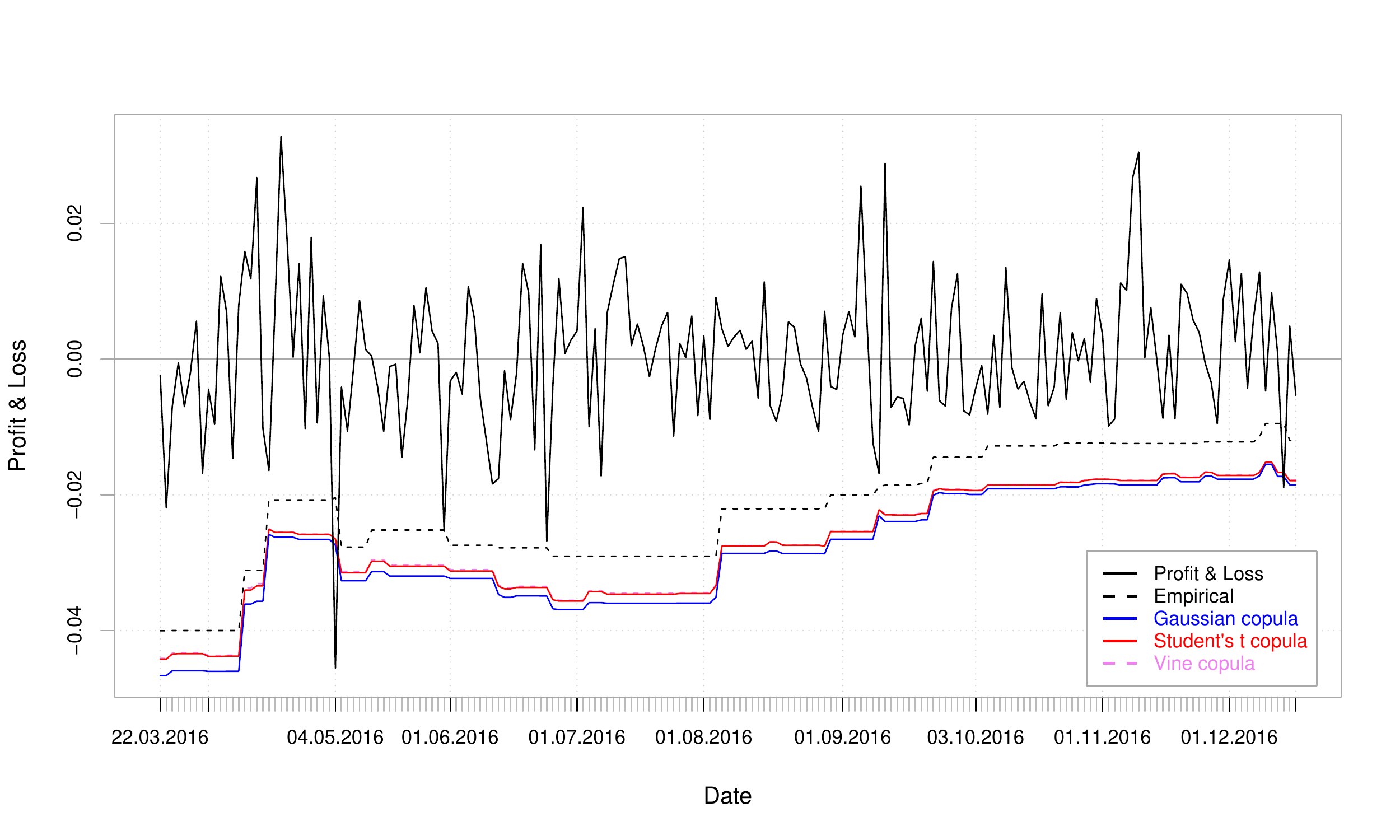}
  \caption{Profit \& Loss (in black), $CVaR$ curve (at the 95\% level) obtained empirically and estimated by Gaussian\,/\,Student's~$t$\,/\,Vine copula}
  \label{ES-curve}
\end{figure}

Fig.~\ref{VaR-curve} and \ref{ES-curve} show the dynamics of Profit \& Loss series and the movement of 95\%-level $VaR$ and $CVaR$ respectively. $VaR$ and $CVaR$ curves simulated by copula models are lower than a historical value. The $p$-values of the Kupiec's $VaR$ test \cite{Kupiec95} for all proposed copula models are greater than the critical level $\alpha = 0.05$. All three models have superior 
prediction ability than usual empirical method.  

\label{Kupiec}

\section{Conclusion}

In the paper we have examined copula models as an approach to describe  multivariate dependence in time series to compare the financial risk measures in framework of portfolio management.
Throughout this paper we make use of 
time series of closing daily prices of stock futures 
for companies: MMC Norilsk Nickel PJSC (GMKR), Gazprom PJSC
(GAZP), Sberbank PJSC (SBRF) respectively, and the future on RTS
index (RTS) during a period of one year from to 17 December, 2015 to
16 Devember, 2016
(four series composed of 253 observations).

We used the two-stage parametric maximum likelihood estimation approach and two copula models -- multivariate
copula and regular vine (R-vine) copula -- have been fitted. We choose the four-parameter hyperbolic, stable and Meixner distributions as the three possible forms of marginal distributions.

Multivariate finance data has been analyzed using three possible distribution models, and it is observed that the all proposed models provide a quite satisfactorily fit to the above data set. According to results of the Cramér–von Mises test, we take the Mexiner distribution as the marginal for all underlying assets.

Using the "Inversion of Kendall’s tau" method we estimated the parameters for multivariate Gaussian, Student’s~$t$, and R-vine copulas and then we executed the parametric
bootstrap-based goodness-of-fit test. The estimated
R-vine copula is given by six pairs based on the Survival Gumbel,
the Survival Clayton, the Survival Clayton-Gumbel, the Joe-Clayton and the Independence copulas.
By analyzing the full Gaussian, Student’s~$t$ and R-vine copulas for the risk management of a portfolio of futures, we find that the impact of copula selection is large. The tests do not reject the R-vine copula, but do reject the Gaussian and  Student’s~$t$ copula.

We proposed the algorithm based on Monte-Carlo simulation of pseudo-observations to compute the $VaR$ and $CVaR$ of the optimal portfolio.  $VaR$ and $CVaR$ curves simulated by copula models are lower than a historical value. All three copula models have superior 
prediction ability than a usual empirical method.

The further research of our study can be continued
in the following directions.
At first, we will compute the associated 95\% confidence intervals of the the unknown parameters. 
At second, we will also investigate the usage the set of spanning trees (for instance, top-10 spanning trees) instead of using just the one maximum spanning tree in the selection of the R-vine array structure.
At third, we will show how one can detect and exclude arbitrage opportunity and avoid these in method of generation scenario trees using copula models. Thereby, we will be consistent with financial asset pricing theory.



\end{document}